\documentclass[%
aps,prl,reprint,superscriptaddress,
nobibnotes,
amsmath,amssymb, aps, 
]{revtex4-2}

\usepackage{graphicx}
\usepackage{dcolumn}
\usepackage{bm}
\usepackage{siunitx}
\usepackage{hyphenat}
\usepackage{flafter}
\usepackage{float}
\usepackage{color,soul}

\begin{document}

\preprint{APS/123-QED}

\title{Spectroscopic Neutron Imaging for Resolving Hydrogen Dynamics \\ Changes in Battery Electrolytes}

\author{E.~R.~Carre\'on~Ruiz}%
\affiliation{Electrochemistry Laboratory (LEC), Paul Scherrer Institut (PSI), 5232 Villigen PSI, Switzerland}

\author{J.~Lee}
\affiliation{Electrochemistry Laboratory (LEC), Paul Scherrer Institut (PSI), 5232 Villigen PSI, Switzerland}
\affiliation{Laboratory for Neutron Scattering and Imaging (LNS), Paul Scherrer Institut (PSI), 5232 Villigen PSI, Switzerland}

\author{J.~I.~M\'arquez~Dami\'an}
\affiliation{Spallation Physics Group, European Spallation Source ERIC, P.O. Box 176, 22100 Lund, Sweden}

\author{M. Strobl}
\affiliation{Laboratory for Neutron Scattering and Imaging (LNS), Paul Scherrer Institut (PSI), 5232 Villigen PSI, Switzerland}
\affiliation{Niels Bohr Institute, University of Copenhagen, Nørregade 10, 1165 Copenhagen, Denmark}

\author{G.~Burca}
\affiliation{STFC-Rutherford Appleton Laboratory, ISIS Facility, Harwell OX11 0QX, UK}
\affiliation{Faculty of Science and Engineering, The University of Manchester, Alan Turing Building, Oxford Road, Manchester, M13 9PL, UK}

\author{R.~Woracek}
\affiliation{European Spallation Source ERIC, P.O. Box 176, 22100 Lund, Sweden}

\author{M.~Cochet}
\affiliation{Electrochemistry Laboratory (LEC), Paul Scherrer Institut (PSI), 5232 Villigen PSI, Switzerland}

\author{M.-O.~Ebert}
\affiliation{Department of Chemistry and Applied Biosciences, ETH Zurich, Vladimir-Prelog-Weg 1-5, 8093 Zurich, Switzerland}

\author{L.~H\"oltschi}
\affiliation{Electrochemistry Laboratory (LEC), Paul Scherrer Institut (PSI), 5232 Villigen PSI, Switzerland}

\author{P.~M.~Kadletz}
\affiliation{European Spallation Source ERIC, P.O. Box 176, 22100 Lund, Sweden}

\author{A.~S.~Tremsin}
\affiliation{Space Sciences Laboratory, University of California, Berkeley, CA 94720, USA}

\author{E.~Winter}
\affiliation{Electrochemistry Laboratory (LEC), Paul Scherrer Institut (PSI), 5232 Villigen PSI, Switzerland}

\author{M.~Zlobinski}
\author{L.~Gubler}
\affiliation{Electrochemistry Laboratory (LEC), Paul Scherrer Institut (PSI), 5232 Villigen PSI, Switzerland}

\author{P.~Boillat}
\affiliation{Electrochemistry Laboratory (LEC), Paul Scherrer Institut (PSI), 5232 Villigen PSI, Switzerland}
\affiliation{Laboratory for Neutron Scattering and Imaging (LNS), Paul Scherrer Institut (PSI), 5232 Villigen PSI, Switzerland}


\begin{abstract}

We present spectroscopic neutron imaging (SNI), a bridge between imaging and scattering techniques, for the analysis of hydrogenated molecules in lithium-ion cells. The scattering information of CH\textsubscript{n}–based organic solvents and electrolytes was mapped in two-dimensional space by investigating the wavelength-dependent property of hydrogen atoms through time-of-flight imaging. Our investigation demonstrates a novel approach to detect physical and chemical changes in hydrogenated liquids, which extends, but not limits, the use of SNI to relevant applications in electrochemical devices, e.g., the study of electrolytes in Li-ion batteries.

\end{abstract}

\maketitle

\textit{Introduction.\textbf{---}} Capabilities of neutron imaging (NI) have been significantly expanded with the advent of energy-resolved neutron imaging, which has become an essential tool for non-destructive material characterization \cite{kardjilov_neutron_2017, woracek_diffraction_2018, carminati_bragg-edge_2020, strickland_2d_2020, dabah_time-resolved_2017, shinohara_energy-resolved_2020, tran_spectral_2021}, including studies of electrochemical energy storage and conversion devices \cite{siegwart_distinction_2019, biesdorf_dual_2014, higuchi_pulsed_2021}. The particular advantage of NI to study light elements of technological importance, i.e., hydrogen and lithium, even in \textit{operando} processes has contributed to the improvements of green mobility technologies such as fuel cells \cite{forner-cuenca_advanced_2016, manzi-orezzoli_impact_2020} and lithium-ion batteries \cite{goers_situ_2004,wang_situ_2012} (LIBs).

Electrodes in LIBs have been extensively studied by conventional (white\hyp{beam}) NI \cite{owejan_direct_2012, nie_probing_2019, zhou_probing_2016, siegel_neutron_2011, riley_situ_2010}, and energy\hyp{resolved} NI has been effective in determining the state\hyp{of-charge} via Bragg\hyp{edge} analysis \cite{lacatusu_multimodal_2021, kamiyama_structural_2017}. Physical or chemical changes in the electrolytes, on the other hand, are typically studied using spectroscopic techniques including Raman, nuclear magnetic resonance, and Fourier\hyp{transform} infrared spectroscopies \cite{gachot_deciphering_2008,jafta_ion_2018, shi_identification_2014,nowak_review-chemical_2015, friesen_impact_2016, vortmann-westhoven_where_2017, pabst_molecular_2017,santner_-situ_2004, ramesh_ftir_2007, peng_multinuclear_2018, hayamizu_temperature_2012, hayamizu_pulse-gradient_1999, oldiges_understanding_2018, laruelle_identification_2004}. However, these techniques usually require specifically designed cells or involve complicated procedures to extract the electrolytes. Therefore, it is common to perform \textit{ex situ} mimicking \textit{operando} conditions.

Neutron instrumentation development (detection systems \cite{tremsin_unique_2020, miller_high-resolution_2020, trtik_progress_2016, borges_event_2018, losko_new_2021, woracek_spatially_2019}, methods \cite{strobl_polarization_2019, woracek_diffraction_2018, strobl_small_2017}) narrowed the gap \cite{lehmann_new_2017, boillat_neutron_2017} between information accessible by imaging and scattering techniques \cite{strobl_scope_2015, woracek_test_2016, ramadhan_characterization_2019}, which lead to the designed of new beamlines (e.g., ODIN, ESS \cite{strobl_scope_2015}) to exploit new opportunities specially in the NI field. In this work, we demonstrate that a study of battery electrolytes (specifically liquid organic solvents) has become feasible with spectral techniques as time\hyp{of-flight} (ToF) NI, and we present an approach to extend it towards a spectroscopic method to resolve physical and chemical changes by analyzing the H total attenuation cross-section. We aim to develop a methodology to study hydrogenated molecules in lithium-ion battery electrolytes \textit{in situ} and even \textit{operando}. 

\textit{Bridging the gap between neutron imaging and spectroscopy.\textbf{---}} NI techniques are based on neutron interaction with matter, which is described by the Beer\hyp{Lambert} law, $T = e^{-\Sigma\delta}$, where \textit{T} is the transmission, $\delta$ is the path length, and $\Sigma$ is the attenuation coefficient of the material. While conventional NI is performed with a polychromatic neutron beam (no energy selection), energy\hyp{resolved} NI is performed with monochromators or the ToF method, which requires the use of pulsed sources or chopper disks that facilitate to spread neutrons in time according to their energy. The speed of neutrons is proportional to the square root of their energy ($E$), whose equivalence to the wavelength ($\lambda$) is given by the de Broglie relation, $\lambda = \textit{h}/\sqrt{2m_{n}E}$ \cite{dianoux_neutron_2003}, where \textit{h} is the Planck constant,  and \textit{m\textsubscript{n}} is the neutron mass. Attenuation coefficients are related to the number density, \textit{N\textsubscript{i}}, of nuclei in the sample and the neutron total cross\hyp{section} $\sigma\textsubscript{T}\textit(\lambda)$. $\sigma\textsubscript{T}\textit(\lambda)$ is the sum of the absorption $\sigma\textsubscript{a}\textit(\lambda)$, coherent $\sigma\textsubscript{coh}\textit(\lambda)$, and incoherent $\sigma\textsubscript{inc}\textit(\lambda)$ scattering cross\hyp{sections}, and they are energy\hyp{dependent} for each atom \cite{adair_neutron_1950}, compound \cite{herdade_neutron_1973, armstrong_energy-dependent_1965, lee_neutron_2020}, and intramolecular bonds \cite{romanelli_thermal_2021}.

In material science and condensed matter research, dynamic processes are studied by inelastic neutron scattering (INS) and quasi\hyp{elastic} neutron scattering (QENS), which take advantage of the large $\sigma\textsubscript{inc}\textit(\lambda)$ contribution of \textsuperscript{1}H \cite{sears_neutron_1992, triolo_quasielastic_2003, lautner_dynamic_2017, embs_introduction_2010}. The quantity measured in QENS experiments is the double differential scattering cross-section (DDSCS), $\frac{\partial^{2}\sigma }{\partial \Omega \partial \lambda'} = \frac{k_{f}}{k_{i}}\left\{ \frac{\sigma_{coh}}{4\pi} S_{coh}(\alpha,\beta)  + \frac{\sigma_{inc}}{4\pi} S_{inc}(\alpha,\beta)\right\}$, where $k\textsubscript{i}$ is the wavenumber of the incident neutrons, k\textsubscript{f} is the wavenumber of the scattered neutrons, \textit{S} is the coherent and incoherent coefficient in the thermal scattering law \cite{embs_introduction_2010, romanelli_thermal_2021, nuclear_energy_agency_nea_thermal_2020}, and $\alpha$ and $\beta$ are momentum and energy transfer variables, respectively \cite{ramic_thermal_2018, nuclear_energy_agency_nea_thermal_2020}. The scattering cross-section,  $\sigma\textsubscript{sct}\textit(\lambda)$, and therefore the \textit{N\textsubscript{i}} of nuclei  ($\sigma\textsubscript{coh}\textit(\lambda)$ + $\sigma\textsubscript{inc}\textit(\lambda)$), is calculated from the DDSCS, $\sigma_{sct}\textit(\lambda) = \int_{}\int{}\frac{\partial^{2}\sigma }{\partial \Omega \partial \lambda'}  d\lambda' d\Omega$ \cite{nuclear_energy_agency_nea_thermal_2020}.

In protonated organic molecules, such as solvents used in Li-ion cells, the main contribution to the signal derives from the \textsuperscript{1}H incoherent scattering cross-section (coherent cross-section: 1.76 barns and incoherent cross-section: 80.26 barns) \cite{faraone_quasielastic_2001,bee_quasielastic_1988, ramic_thermal_2018}. The remaining elements, i.e., carbon and oxygen, have a negligible incoherent scattering cross-section (0.0010 barns and 0.0008 barns, respectively), and in this study, their contributions were corrected by subtracting the corresponding free atom scattering cross section (5.551 barns and 4.232 barns, respectively) \cite{capelli_effective_2019} to obtain just the $\mathrm{\sigma}_{T}^{mol}$ of the molecule, i.e., $\mathrm{\sigma}_{T}^{mol} = A\mathrm{\sigma}_{T}^{C} + B\mathrm{\sigma}_{T}^{O} + C\mathrm{\sigma}_{T}^{H}$, where A, B, and C are the number of atoms in the molecule, respectively. This process is similar to the average functional group approximation, described by Romanelli et al. \cite{romanelli_thermal_2021}. With these considerations, a transmission image, \textit{T}, contains the elastic and inelastic components of the incoherent scattering process ($\mathrm{S}_{inc}^{el}(\alpha) + \mathrm{S}_{inc}^{inel} (\alpha,\beta)$). The elastic incoherent part carries information about the quantity and type of scattering atoms. On the other hand, the inelastic component provides information on the motion of these atoms, due to molecular vibration or diffusion processes. In this way, and by applying the Beer-Lambert law to calculate the \textsuperscript{1}H total neutron attenuation cross-section, $\mathrm{\sigma}_{T}^{H}(\lambda)$, we exploit for the first time that transmission images obtained by ToF-NI provide spatially resolved information on the chemical and physical properties of organic materials. The fully non-invasive aspect of neutron imaging allows investigating electrolyte degradation and/or physical changes within real batteries (i.e., while confined in metallic casings), which are crucial to understanding performance limitations in LIBs.

\textit{Experimental.\textbf{---}} Two sets of experiments were performed at two neutron facilities: a reactor neutron source (V20 ESS test-beamline at HZB in Berlin) \cite{woracek_test_2016}, where the physical pulses were generated with a chopper system, and a pulsed spallation source (IMAT beamline of ISIS RAL in the UK)\cite{kockelmann_time--flight_2018}. The first experiment (V20) was performed with coin cells filled with organic solvents and electrolytes in a through-plane setup at a wavelength range of 1.50{\AA} to 9.6{\AA}. The total exposure time for each sample batch (two coin cells at the same time) was fixed to two hours with 5-min ToF imaging cycles. The second experiment (IMAT) was conducted with two wavelength ranges, 0.7-6.5{\AA}, and 3.5-9.7{\AA}. The total exposure time was fixed to four hours and one hour, respectively, per sample batch (four cuvettes at the same time) with 5-min ToF imaging cycles. To probe the transmission of organic solvents and electrolytes and to calculate the $\mathrm{\sigma}_{T}^{H}(\lambda)$, we used a stainless steel holder including 16 cuvettes 3mm thickness to optimize the utilization of the limited field-of-view (FoV) in the detector. A micro-channel plate (MCP) detector \cite{tremsin_high_2013} with a 28x28mm$^{2}$ FoV and $\SI{55}{\micro\metre}$ pixel size in normal transmission geometry was used for both measurements. 

The samples consist of pure organic solvents from BASF and Sigma Aldrich: ethylene carbonate (EC), dimethyl carbonate (DMC), and diethyl carbonate (DEC). Organic binary mixtures: EC-DMC 1:1 w/w, EC-DEC 1:1 v/v and EC-DEC 3:7 v/v. Electrolytes: LP30 (1M LiPF\textsubscript {6} in EC-DMC 1:1 w/w), LP40 (1M LiPF\textsubscript{6} in EC-DEC 1:1 v/v), LP47 (1M LiPF\textsubscript{6} in EC-DEC 3:7 w/w). The CH\textsubscript{n} bonds included in each organic molecule are: two CH\textsubscript{2} bonds in EC, two CH\textsubscript{3} bonds in DMC, and one pair of CH\textsubscript{2} and CH\textsubscript{3} bonds in DEC. Comparison between IMAT and V20 cross-section results can be found in the supplementary material section (appendix-A).

\begin{figure}
\centering \includegraphics[width=\linewidth]{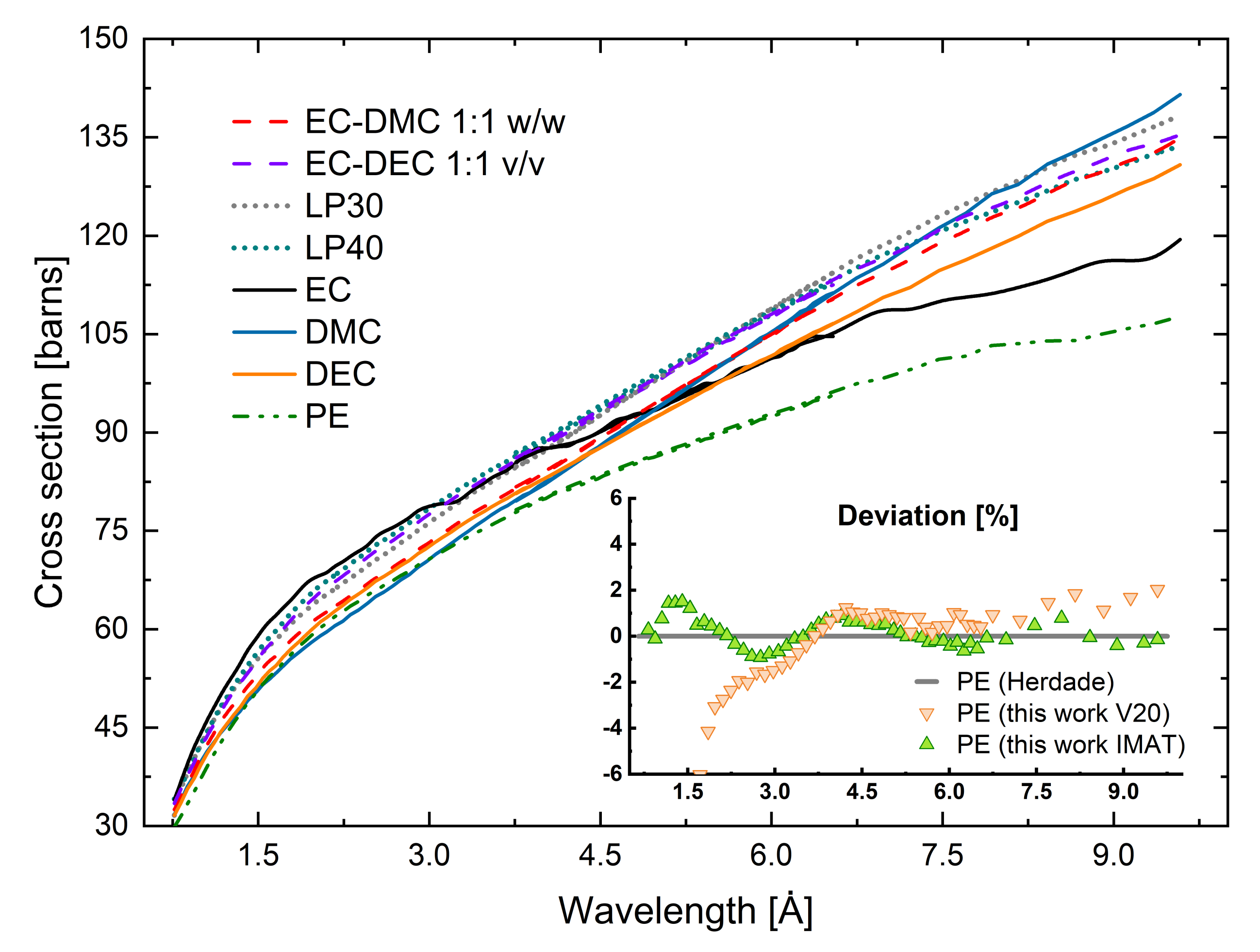}
\caption{\label{fig:1} Calculated $\mathrm{\sigma}_{T}^{H}(\lambda)$ in organic solvents, organic binary mixtures, and electrolytes (IMAT beamline). Cross-section dependence on the nature of the molecule to which H is bonded. Down right graph: percentage deviation between calculations of $\mathrm{\sigma}_{T}^{H}(\lambda)$ in polyethylene in our experiments, as a reference material, compared to literature data. \cite{herdade_neutron_1973,lee_neutron_2020}.}
\end{figure}

\textit{Results and discussion.\textbf{---}} The calculation of the $\mathrm{\sigma}_{T}^{H}(\lambda)$ from transmission images in ToF-NI experiments (Fig.\ref{fig:1}) revealed variations in the \textsuperscript{1}H scattering behavior in different compounds. A good agreement was found between our measurements of \textsuperscript{1}H in polyethylene (PE), as reference material, and literature data \cite{herdade_neutron_1973, lee_neutron_2020} with an error below 1.5\% (at the IMAT) shown in Fig.\ref{fig:1} down right. The natural fluctuation of $\mathrm{\sigma}_{T}^{H}(\lambda)$ results in a wavelength-dependent distinction between organic materials, based on the atomic bonds, intermolecular interactions, and the overall dynamic characteristics of each molecule including diffusion.

\begin{figure}
\centering\includegraphics[width=\linewidth]{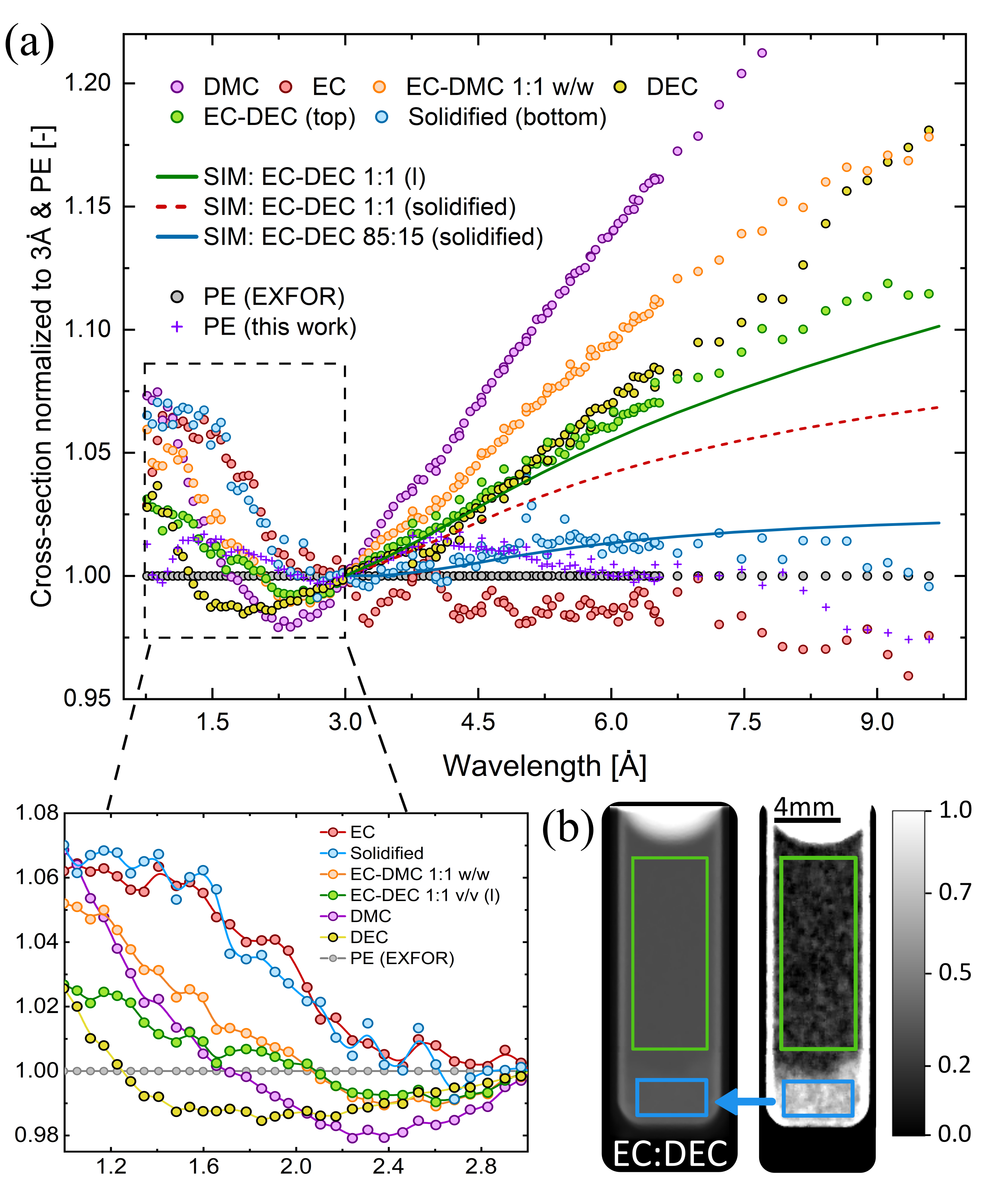}
\caption{\label{fig:2} Partial solidification of an organic binary mixture. (a) \textsuperscript{1}H cross-sections normalized to PE at 3{\AA}: Liquid binary mixtures (EC-DEC 1:1 v/v and EC-DMC 1:1 w/w) and pure organic solvents (EC and DMC). The blue and green curves correspond to the simulation of the solid deposit and the liquid organic binary solvent, respectively, both in the same cuvette. Region 1$<$$\lambda$$<$3{\AA} shows differences among solvents according to their overall molecular vibration. The fine variations depend on the aggregation state of the sample and chemical structure (H fraction in CH\textsubscript{n} groups).} (b) Neutron radiograph at 3{\AA}. Green and blue regions were used in the calculation of the $\mathrm{\sigma}_{T}^{H}(\lambda)$ for each component.
\end{figure}

The curves shown in Fig.\ref{fig:1} are a rater complex convolution of molecular dynamic features that are not easily discerned, which is the fundamental reason to use simulations. In our approach, we associate simulations and measurements to identify the contributions that cause the fine variations in the curves. Therefore, to accentuate said divergence, the $\mathrm{\sigma}_{T}^{H}(\lambda)$ of each sample were normalized to that of PE reference and divided by its cross-section value at 3{\AA} (Fig.\ref{fig:2}-a). At short $\lambda$ ($<$3{\AA}), H cross-sections provide information of the overall vibrational features of the molecule. Molecules with the same vibrational modes overlap, i.e., CH\textsubscript{2}-rich molecules on top, molecules with a combination of CH\textsubscript{2}-rich and CH\textsubscript{3} molecules in the middle, pure molecules containing a balanced combination of CH\textsubscript{2} and CH\textsubscript{3} at the bottom, and molecules containing pure CH\textsubscript{3} diverging along the region. Thus, a separation between CH\textsubscript{n} (especially between CH\textsubscript{2} and  CH\textsubscript{3}) functional groups can be observed (Fig.\ref{fig:2}-a inset). A similar analysis was performed at long $\lambda$ ($>$3{\AA}), where the change is even more substantial, and additional information about a concentration change of a molecule can be obtained. In both short and long $\lambda$ regions, information about the aggregation state of the sample can be found. The evidence of this particular behavior along the spectrum originated from the observation of a partial solidification at experiment temperature (290K) (Fig.\ref{fig:2}-b) in an organic binary solvent (EC-DEC 1:1 v/v), which is the base of the LP40 electrolyte (1M LiPF\textsubscript{6} in EC-DEC 1:1 v/v). The radiograph revealed a deposit at the bottom of the cuvette with a different transmission signal to that of the top part. The contrast variation is attributed either to the presence of solid EC or a solid mixture containing high EC concentration. The phase change and the solid aggregation state of the deposit were confirmed considering the curve has a comparable behavior as EC (melting point 312K) in both the vibrational and diffusion regions.

The significant difference of $\mathrm{\sigma}_{T}^{H}(\lambda)$ between the liquid and solid curves in the diffusion region are due to the solidification (phase change effect), a change of composition (mixture EC-DEC with an EC enrichment, a concentration change effect), or both. A $\mathrm{\sigma}_{T}^{H}(\lambda)$ simulation (appendix-B) of this solvent (dashed red curve in Fig.\ref{fig:2}-a), where the self-diffusion coefficient is set to zero, i.e., assuming solidification of EC-DEC 1:1 v/v without changing concentrations, indicates that a reduction of the slope in the diffusion region of the material is expected. However, it is rather small compared to our observations. From the phase diagram for the mixture EC-DEC \cite{ding_liquidsolid_2001}, the co-existence of a solid EC phase with an EC-depleted liquid phase is expected at 290K. Therefore, a simulation of this component in solid state with a higher EC concentration (85:15) was performed. The blue curve in Fig.\ref{fig:2}-a mimics the change of concentration and shows a better fit to the slope reduction observed experimentally indicating both the change of concentration and the phase transition to solid state of the sample. The major part of the slope reduction in our measurements is caused by the separation of an EC-enriched solid deposit, which was accumulated at the bottom of the sample.

\begin{figure}[H]
\centering\includegraphics[width=\linewidth]{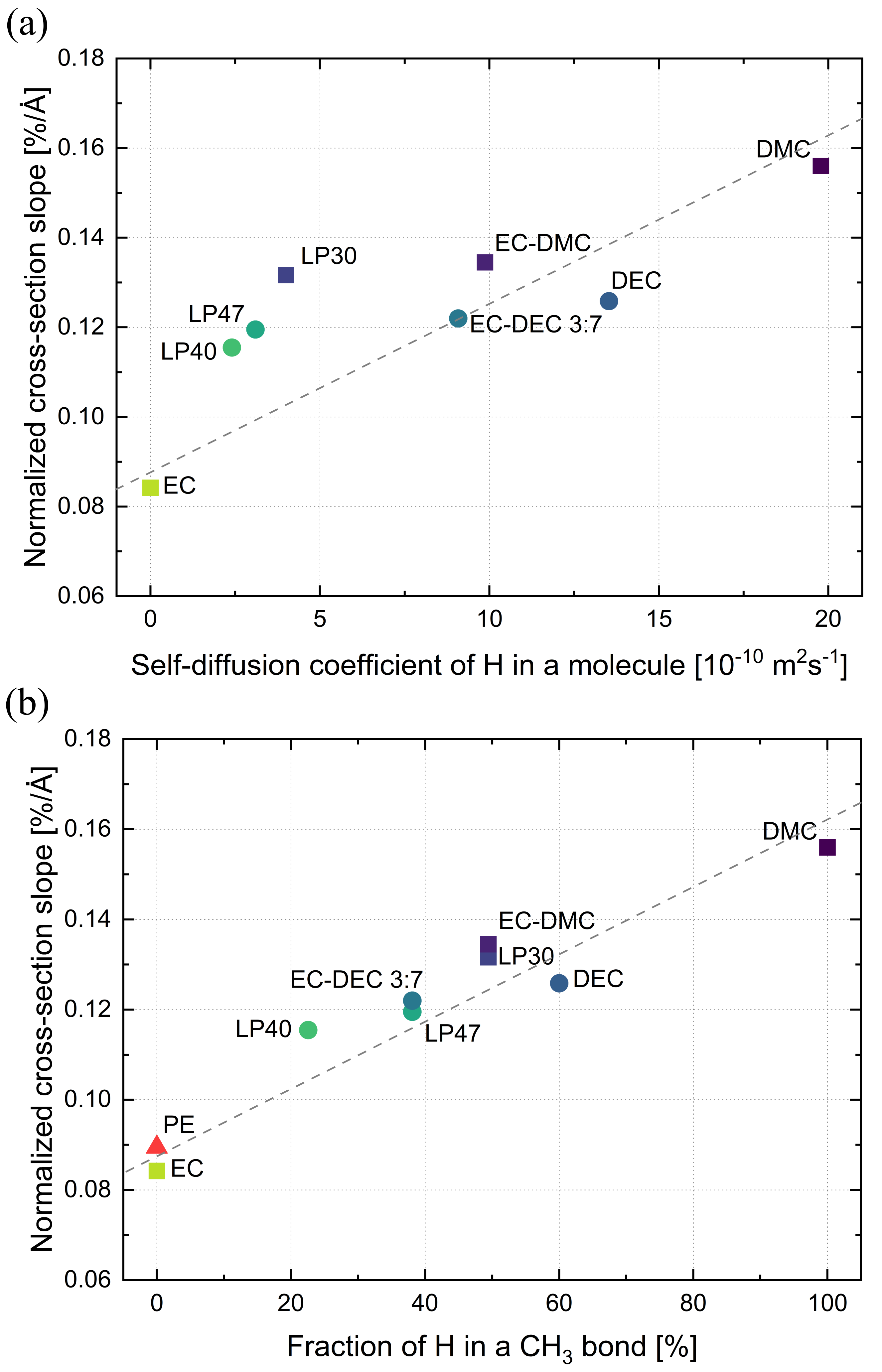}
\caption{\label{fig:3} Relationship between the $\mathrm{\sigma}_{T}^{H}(\lambda)$ normalized slope and (a) the measured \textsuperscript{1}H self-diffusion coefficients of pure organic solvents, organic binary mixtures, and electrolytes. (b) the fraction of H in CH\textsubscript{3} bonds in each molecule. Squared markers represent solvents whose individual molecules contain pure CH\textsubscript{2} (EC), CH\textsubscript{3} (DMC) or a pure combination of CH\textsubscript{2} and CH\textsubscript{3} (EC-DMC) functional groups , circular markers represent individual molecules with both CH\textsubscript{2} and CH\textsubscript{3} functional groups, such as DEC.}
\end{figure}

To analyze the relationship between $\mathrm{\sigma}_{T}^{H}(\lambda)$ at long $\lambda$ and the properties of the molecule, like viscosity, the slope of the normalized cross-sections in the diffusion region ($\lambda >$3{\AA}) was calculated. In addition, the H self-diffusion coefficients of all samples were measured at 290K by NMR spectroscopy, by which the binary diffusion coefficients of organic binary mixtures and electrolytes were calculated. Fig.\ref{fig:3}-a shows, as a first hypothesis, the relationship between the $\mathrm{\sigma}_{T}^{H}(\lambda)$ slope values at long $\lambda$ with the diffusion coefficients. One should expect a distinction in the slope of molecules with substantially different mobility, i.e., the binary mixture EC-DEC 3:7 and the electrolyte LP47. However, this is true only for pure solvents (EC, DEC, and DMC). This fact suggests that the $\mathrm{\sigma}_{T}^{H}(\lambda)$ slope is not that much influenced by the H self-diffusion, but rather by the type of chemical bond, as suggested by our simulations. Particularly, the slope takes into account the nature and fraction of CH\textsubscript{2} and CH\textsubscript{3} in each molecule, i.e., DEC contains a combination of both CH\textsubscript{2} and CH\textsubscript{3} functional groups (circular markers) in contrast to pure and combinations of pure CH\textsubscript{n} functional groups (squared markers). A confirmation of this hypothesis is observed in Fig.\ref{fig:3}-b, where the molar fraction of H in CH\textsubscript{3} bonds was calculated for each sample. Here, a stronger correlation between the normalized $\mathrm{\sigma}_{T}^{H}(\lambda)$ slope and the fraction of H in CH\textsubscript{3} functional groups is shown.

The slope at long $\lambda$ due to relatively small molecules containing only CH\textsubscript{3} bonds, such as DMC, provide a higher value of $\mathrm{\sigma}_{T}^{H}(\lambda)$ than that of larger molecules containing both CH\textsubscript{3} and CH\textsubscript{2} bonds, such as DEC. Romanelli et al. \cite{romanelli_thermal_2021} observed a similar behavior in the same energy range when determining  $\sigma_{T}(\lambda)$ of amino acids. They explain that the higher value of $\sigma_{T}(\lambda)$ in CH\textsubscript{3} bonds is due to low-energy excitation and de-excitations in its vibrational density of states (VDoS), which fits within numerical accuracy for $\mathrm{\sigma}_{T}^{H}(\lambda)$ in the case of CH, CH\textsubscript{2}, and CH\textsubscript{3} molecules.

The $\mathrm{\sigma}_{T}^{H}(\lambda)$ slope is a simple approach to investigate changes in electrolytes. Still, examining of the full spectra (Fig. \ref{fig:2}) provides more opportunities of identifying the changes, e.g., while the difference of slopes at long $\lambda$ between DEC and EC-DEC 3:7, are imperceptible, the information provided by the vibrational region (low $\lambda$) shows a considerable  contrasting behavior, due to unique features of each molecule.

The substantial signal change in $\mathrm{\sigma}_{T}^{H}(\lambda)$ due to physical changes and intrinsic molecular dynamics perceived along the thermal-to-cold neutron energies (0.6{\AA} to 9{\AA}) constitutes a fundamental strength of the method in addition to its spatially resolved nature. The results obtained in this experiment demonstrate the use of ToF-NI as a spectroscopic method to track phase changes and to differentiate specific organic molecules based on the nature of their bonds. These attributes are crucial to applications in material science and especially in the investigation of electrolyte degradation in LIBs, where the direct study of physico-chemical reactions occurring inside metal casings would otherwise be impossible.

\textit{Conclusion.\textbf{---}} We demonstrate the use of ToF-NI to spatially resolve physico-chemical changes and to obtain spectroscopic information in organic solvents used in LIBs in a non-invasive manner. The images resulting from this technique contain the necessary information to track phase and concentration changes, and to differentiate molecules according to their structure and chemistry. 

The wavelength-dependent \textsuperscript{1}H total cross-sections provide insights to vibrational and diffusion dynamics of organic materials, and their dependence on both: the aggregation state (solid or liquid) and the type of CH\textsubscript{n} functional groups in which H in bound (e.g., CH\textsubscript{2}, CH\textsubscript{3}, and combinations). However, based on our measurements and simulations, the latter (type of CH\textsubscript{n} groups) has a higher impact than the aggregation state.

The variations of the wavelength dependent \textsuperscript{1}H total cross-sections of each material described here demonstrate the unique capabilities of ToF-NI to operate as a spatially resolved spectroscopic method, particularly in environments which cannot be accessed otherwise. Therefore, we foresee the use of ToF-NI in a broader range of applications (e.g., Na-ion and Mg-ion based batteries, redox flow batteries) to study \textit{operando} electrolyte degradation, distribution of heterogeneities, partial solidification, and issues related to mass transport mechanisms. The utilization of this method for LIBs in \textit{in situ} testings and \textit{operando} fast-charging protocols will allow the study of performance losses, since a correlation between physico-chemical and electrochemical performance is viable.

\textit{Acknowledgements.\textbf{---}} We are grateful for fruitful discussions with L. Kondracki, A. Mularczyk, and  D. Scheuble. We acknowledge financial support from the PSI CROSS initiative and scientific collaboration to the IMAT beamline (experiment RB2000122 with DOI: 10.5286/ISIS.E.RB2000122) at ISIS Neutron and Muon Source (UK) of Science and Technology Facilities Council (STFC).

\section{Supplementary material}

\subsection{\label{sec:suppmatA}A. The V20 and the IMAT cross-sections}
Comparisson of \textsuperscript{1}H cross-sections obtained at V20 in HZB and at IMAT in RAL. The results obtained at both institutes agree with an error below 2\% except for the EC sample (5\% at long wavelengths $\lambda >$7{\AA}). The variation is due to the sample preparation process. At V20, coin cells were used. When the liquid EC in the coin cell solidified (solid at room temperature), the thickness along the region of interest (ROI) was inhomogeneous. At IMAT, the thickness was controlled by designing a 16 cuvette rectangular sample holder, which fits the MCP detector FoV in batches of 4 samples per scan.

\begin{figure}[H]
\centering\includegraphics[width=\linewidth]{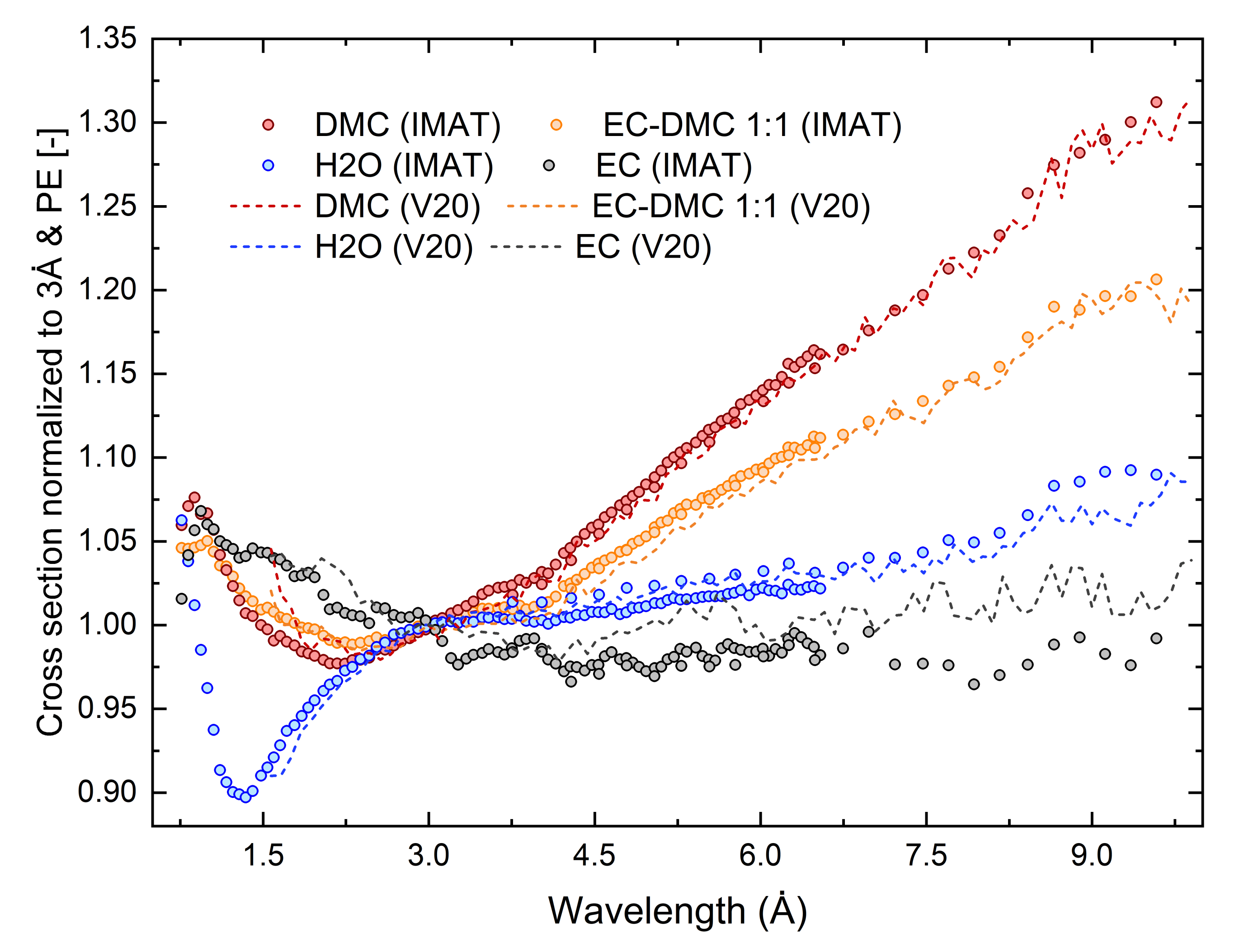}
\caption{\label{fig:supplemental materialA} Calculated \textsuperscript{1}H cross-section of samples measured at V20 and at IMAT. The EC sample measured at V20 (coin cell) was not filled completely and it solidified inhomogeneously. Therefore, the thickness varies along the ROI. X-ray tomography was performed to calculate the average thickness of the EC sample.}
\end{figure}

\subsection{\label{sec:suppmatB}B. The NJOY2016 simulations}
The total cross sections of EC, DEC and binary mixture were modeled using NJOY2016 \cite{macfarlane2017njoy} and the average functional group approximation (AFGA)\cite{romanelli_thermal_2021} model, combined with an Egelstaff-Schofield model for diffusion in the case of liquids. Models were prepared for hydrogen bound in each molecule, and the contribution from minor atoms was computed as a constant, free atom cross section. The phonon spectra of hydrogen bound in solid EC was modeled as 4 hydrogen atoms in CH$_2$ groups, and solid DEC as 4 hydrogen atoms in CH$_2$ groups and 6 hydrogen atoms in CH$_3$ groups. For liquid EC and DEC, the equivalent diffusion mass was computed using a hard-sphere model, following the work by Lin et al. \cite{lin2003two}. This resulted in $M_\text{diff}^\text{EC} = 20.0 M_\text{mol}^\text{EC}$, and $M_\text{diff}^\text{DEC} = 18.9 M_\text{mol}^\text{DEC}$. The phonon ($S_\text{solid}(\alpha, \beta)$) and diffusional ($S_\text{diff}(\alpha\beta)$) contributions to the thermal scattering law were calculated separately, convolved and used to compute the double differential scattering cross section, which was later integrated to obtain the total cross section.

\bibliographystyle{apsrev4-2}
\bibliography{ref}

\end{document}